\documentclass[12pt,twoside]{article}
\usepackage{fleqn,espcrc1,epsfig}
\newcommand{\AmS}{{\protect\the\textfont2
  A\kern-.1667em\lower.5ex\hbox{M}\kern-.125emS}}

\title{ \centering
Production and collective behavior of strange particles in Au\,+\,Au 
collisions at 2--8 AGeV}
\author{ \centering
C. Pinkenburg\address[SUNY]{\vspace{-3mm} \small Depts.~of Physics and Chemistry, State University of New York, Stony Brook}\address[BNL]{\vspace{-3mm} \small Brookhaven National Laboratory, Upton, New York},
        N.N.~Ajitanand\addressmark[SUNY],
        J.M.~Alexander\addressmark[SUNY],
M.Anderson\address[UCD]{\vspace{-3mm} \small University of California, Davis, California},
D.~Best\address[LBL]{\vspace{-3mm} \small Lawrence Berkeley National Laboratory, Berkeley, California},
F.P.~Brady\addressmark[UCD],
T.~Case\addressmark[LBL],
W.~Caskey\addressmark[UCD],
D.~Cebra\addressmark[UCD],
J.L.~Chance\addressmark[UCD],
P.~Chung\addressmark[SUNY],
B.~Cole\address[COLUMBIA]{\vspace{-3mm} \small Columbia University, New York, New York},
K.~Crowe\addressmark[LBL],
A.C.~Das\address[OSU]{\vspace{-3mm} \small Ohio State University, Columbus, Ohio},
J.E.~Draper\addressmark[UCD],
M.L.~Gilkes\addressmark[SUNY],
S.~Gushue\addressmark[BNL],
M.~Heffner\addressmark[UCD],
A.S.~Hirsch\address[PURDUE]{\vspace{-3mm} \small Purdue University, West Lafayette, Indiana},
E.L.~Hjort\addressmark[PURDUE],
L.~Huo\address[HARBIN]{\vspace{-3mm} \small Harbin Institute of Technology, Harbin, P.R.~China},
M.~Justice\address[KENT]{\vspace{-3mm} \small Kent State University, Kent, Ohio},
M.~Kaplan\address[CMU]{\vspace{-3mm} \small Carnegie Mellon University, Pittsburgh, Pennsylvania},
D.~Keane\addressmark[KENT],
J.C.~Kintner\address[STMARY]{\vspace{-3mm} \small St.~Mary's College, Moraga, California},
J.~Klay\addressmark[UCD],
D.~Krofcheck\address[NZ]{\vspace{-3mm} \small University of Auckland, Auckland, New Zealand},
R.A.~Lacey\addressmark[SUNY],
C.~Law\addressmark[SUNY],
J.~Lauret\addressmark[SUNY],
M.A. Lisa\addressmark[OSU],
H.~Liu\addressmark[KENT],
Y.M.~Liu\addressmark[HARBIN],
R.L.~McGrath\addressmark[SUNY],
Z.~Milosevich\addressmark[CMU],
G.~Odyniec\addressmark[LBL],
D.L.~Olson\addressmark[LBL],
S.~Panitkin\addressmark[KENT],
N.T.~Porile\addressmark[PURDUE],
G.~Rai\addressmark[LBL],
H.G.~Ritter\addressmark[LBL],
J.L.~Romero\addressmark[UCD],
R.~Scharenburg\addressmark[PURDUE],
L.S.~Schroeder\addressmark[LBL],
B.~Srivastava\addressmark[PURDUE],
N.T.B.~Stone\addressmark[LBL],
T.J.M.~Symons\addressmark[LBL],
S.~Wang\addressmark[KENT],
J.~Whitfield\addressmark[CMU],
R.~Witt\addressmark[KENT],
L.~Wood\addressmark[UCD],
W.N.~Zhang\addressmark[HARBIN]
}

\begin{document}

% typeset front matter
\maketitle

\normalsize

\begin{abstract}
\small
The E895 experiment at the AGS measured strange particle production and
collective behavior in Au+Au collisions between 2--8 AGeV. The production of
$\Lambda$ Baryons and K$^0$ Mesons as function of energy rises smoothly and 
exhibits a nonlinear impact 
parameter dependence. Neutral and positively charged Kaons
exhibit a strong anti-flow behavior. $\Lambda$ Baryons show a smaller 
flow signal than protons.
\end{abstract}

\vspace{5mm}

The production of strange particles in relativistic heavy ion collisions
is an important probe for high density nuclear matter. Of particular interest
are suggestions that strange particle yields can be used to investigate
the nuclear equation of state \cite{Aichelin1985} or possible signatures 
of a change  of the in-medium 
mass \cite{Weise1996}.
The flow effects of strange particles offer a good probe for their
in-medium potentials \cite{Li1995,Li1996}. 

\begin{figure}[t]
\vspace{-5mm}
\centering\mbox{
\makebox[8.2cm][l]{
\epsfig{file=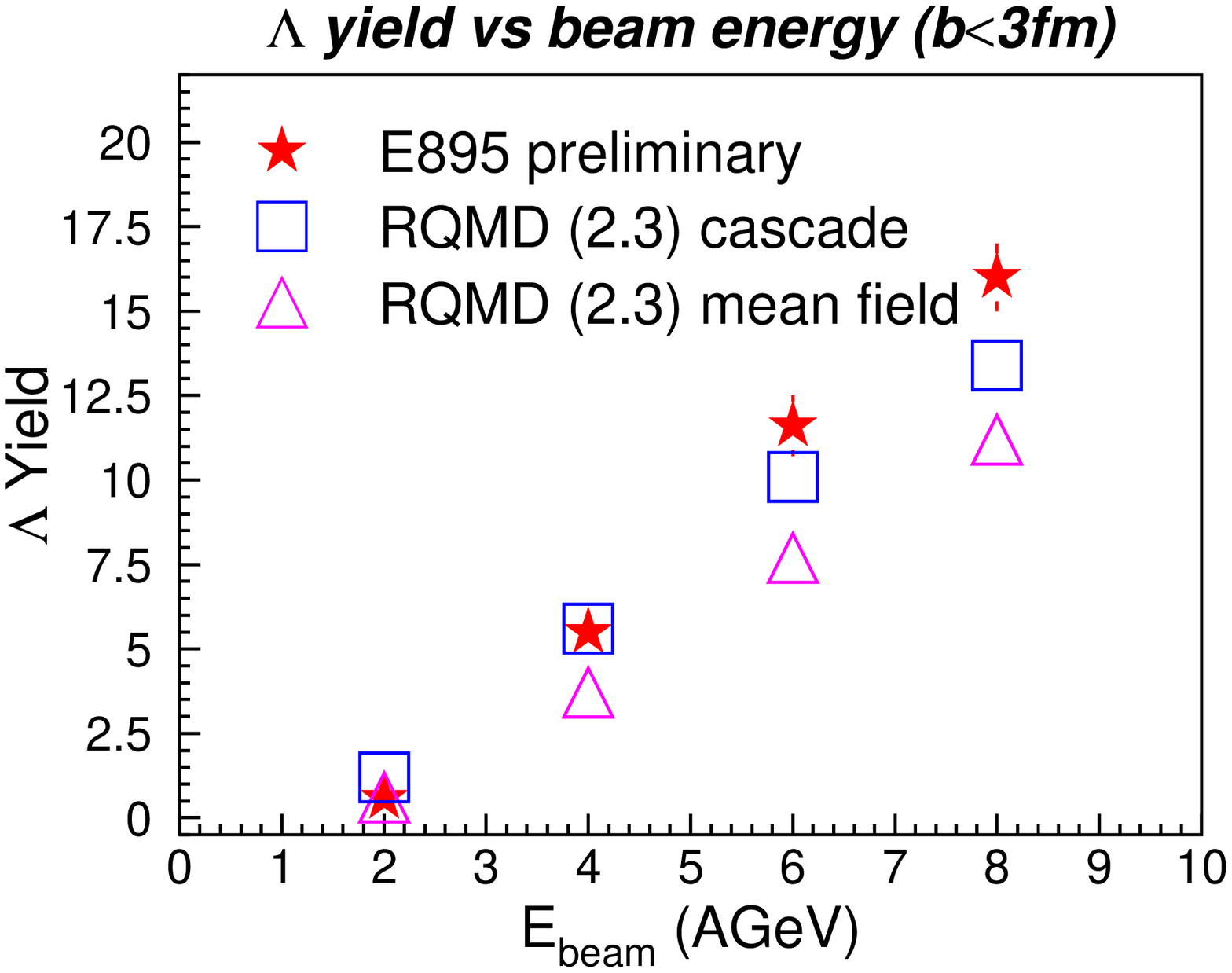,width=8.2cm}
}
\makebox[8cm][r]{
\epsfig{file=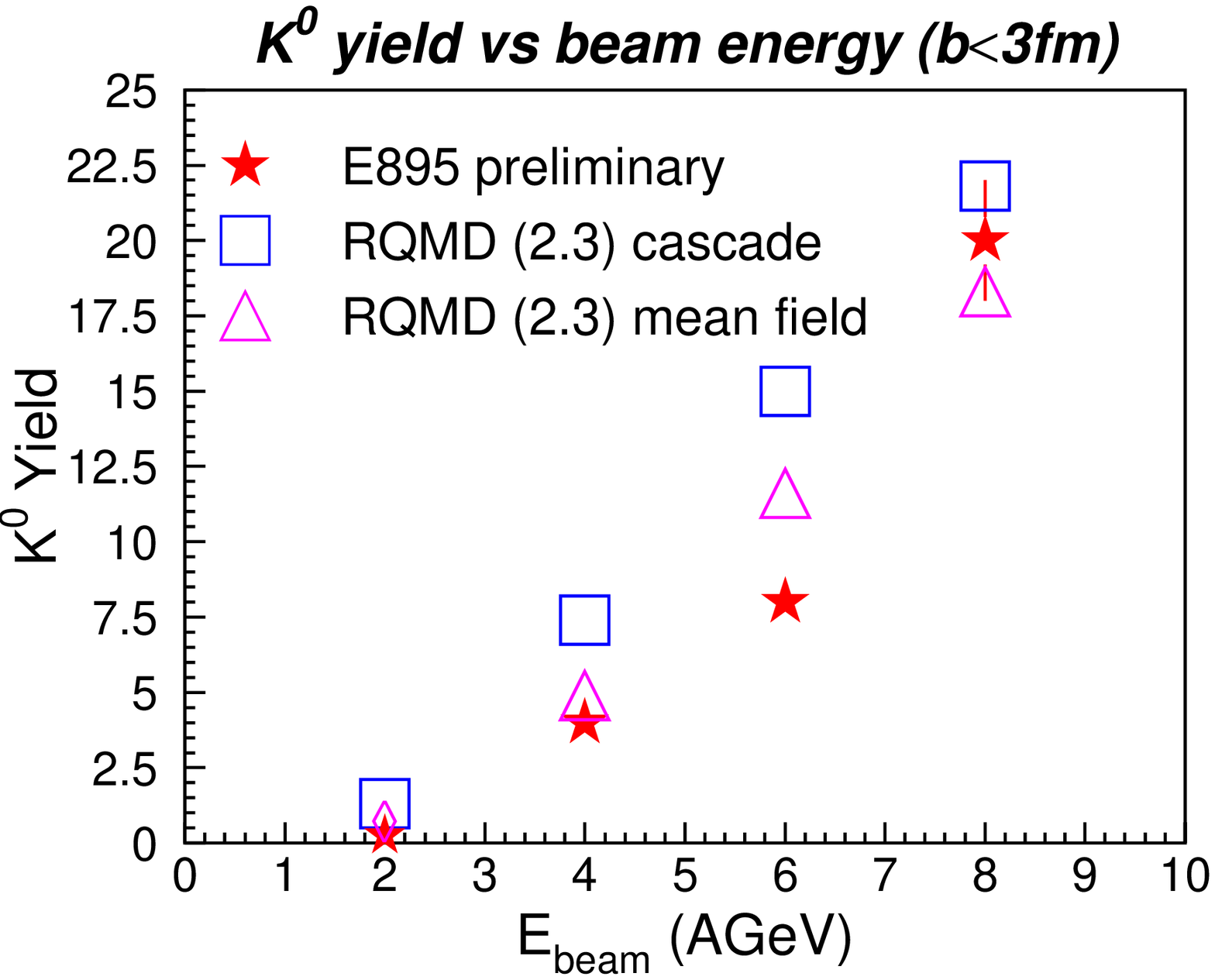,width=8.2cm}
}
}
\vspace{-15mm}
\caption[none]{
\em
\label{fig:v0yields}
\renewcommand{\baselinestretch}{1.}
Mean $\Lambda$ Baryon and K$^0$ Meson multiplicities per event for 
central events. 
}
\vspace{-5mm}
\end{figure}

Motivated by these suggestions the E895 
collaboration
has performed an extensive set of measurements at the Alternating Gradient
Synchrotron (AGS) at the Brookhaven National Laboratory. These measurements
were performed for Au+Au collisions in a beam energy range of 2--8 AGeV. 
Charged reaction products were detected in the EOS Time Projection 
Chamber (TPC) \cite{Rai1990} situated in a uniform magnetic field. The TPC 
provides 
continuous 3D tracking and particle identification for charged particles  
(-1\,$\le$\,Z\,$\le$\,6) with full azimuthal coverage. The method of 
identification via the rigidity of the track and the energy loss in the TPC 
gas leads to ambiguities for certain rigidities, thus restricting the 
identification of charged kaons to low transverse momenta at backward 
rapidities. This restriction does not affect our neutral strange particles 
measurements ($\Lambda$ Baryons and  K$^0_s$ Mesons) which were reconstructed 
from their charged decay products by means of a neural
network \cite{Justice1997,Chung2000,Chung2001}.

\begin{figure}[h]
\vspace{-6mm}
\centering\mbox{
\makebox[8.2cm][l]{
\epsfig{file=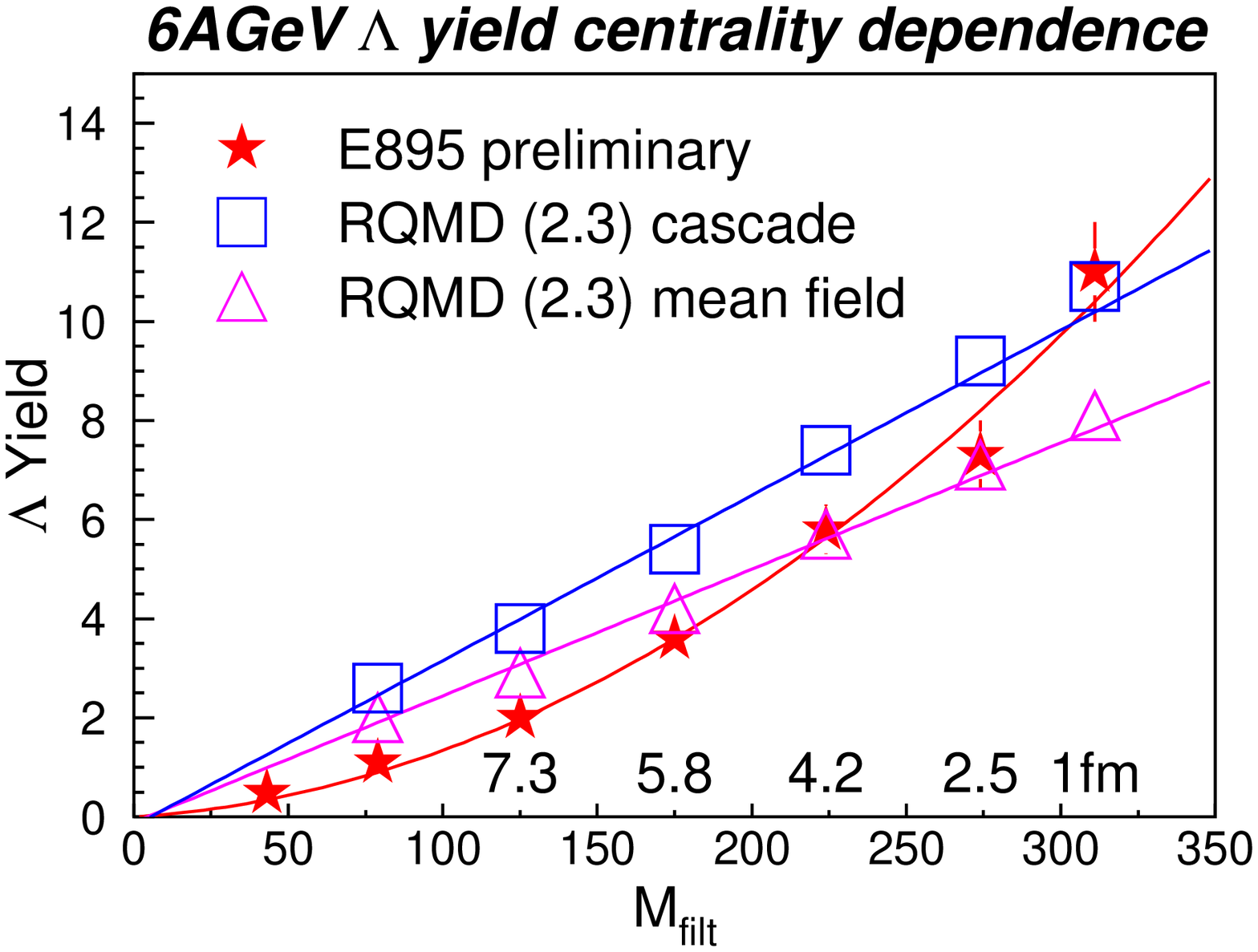,width=8.2cm}
}
\makebox[8cm][r]{
\epsfig{file=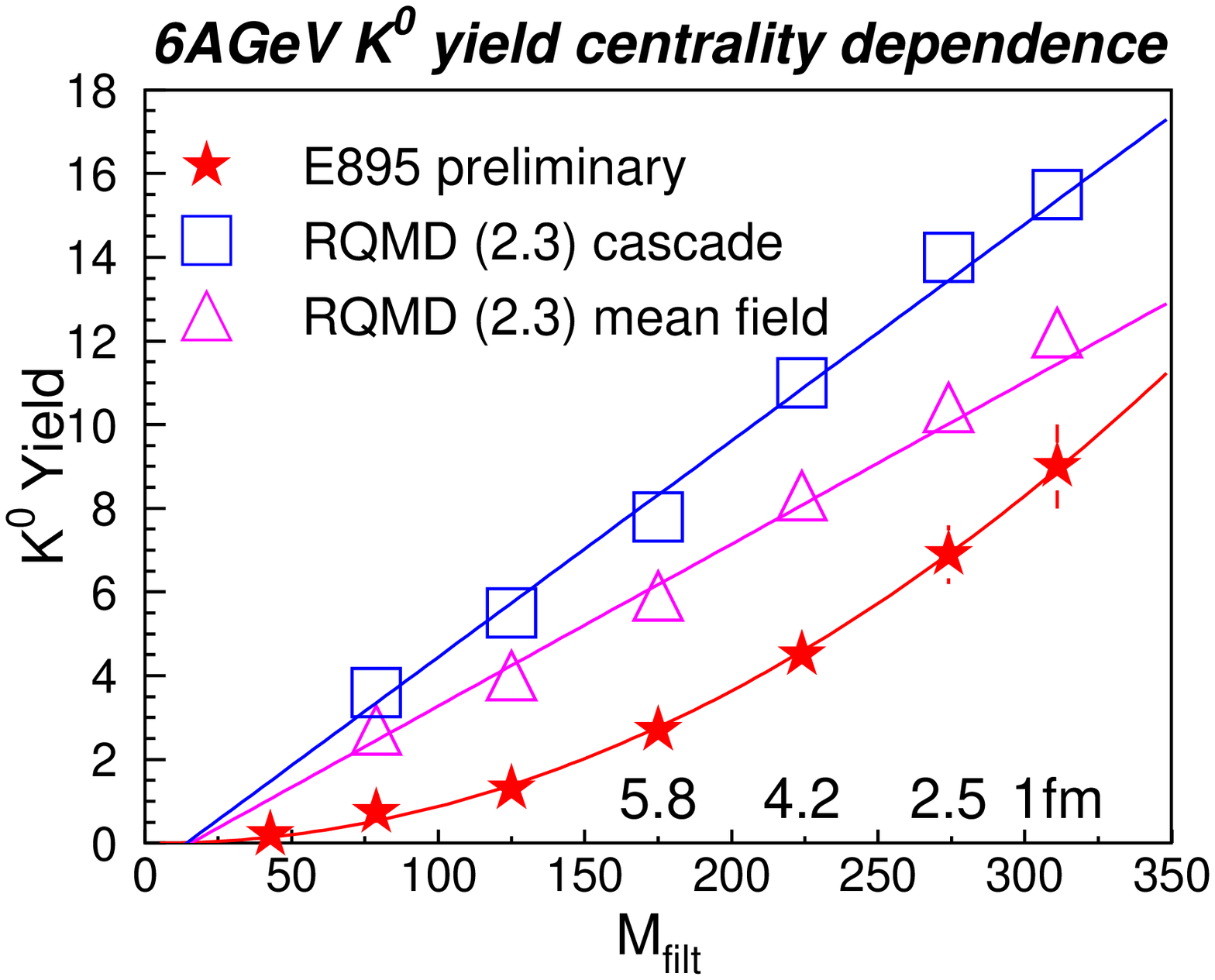,width=8.2cm}
}
}
\vspace{-12mm}
\caption[none]{
\label{fig:v0cent}
\renewcommand{\baselinestretch}{1.} \em
Centrality dependence of the mean multiplicity of $\Lambda$ Baryons
and K$^0$ Mesons in Au+Au collisions at 6 AGeV. The lines are line/parabolic 
fits to guide the eye.
}
\vspace{-5mm}
\end{figure}

Fig.\,\ref{fig:v0yields} shows the mean multiplicities of $\Lambda$ baryons 
and 
K$^0$ Mesons in central events corrected for acceptance and reconstruction 
efficiency.
The yields show a monotonic increase with beam energy and are roughly 
reproduced by the RQMD \cite{Sorge1995} Model. They exhibit no obvious 
features 
which would support a threshold phenomenon. RQMD fails to reproduce the impact 
parameter dependence at 6 AGeV depicted in Fig.\,\ref{fig:v0cent}. The data 
exhibit a strong nonlinear dependence which was previously observed for 
charged Kaons\cite{Ahle1998} for Au\,+\,Au collisions in the same energy range.
 
\begin{figure}[t]
\vspace{-5mm}
\centering\mbox{
\makebox[7.5cm][l]{
\epsfig{file=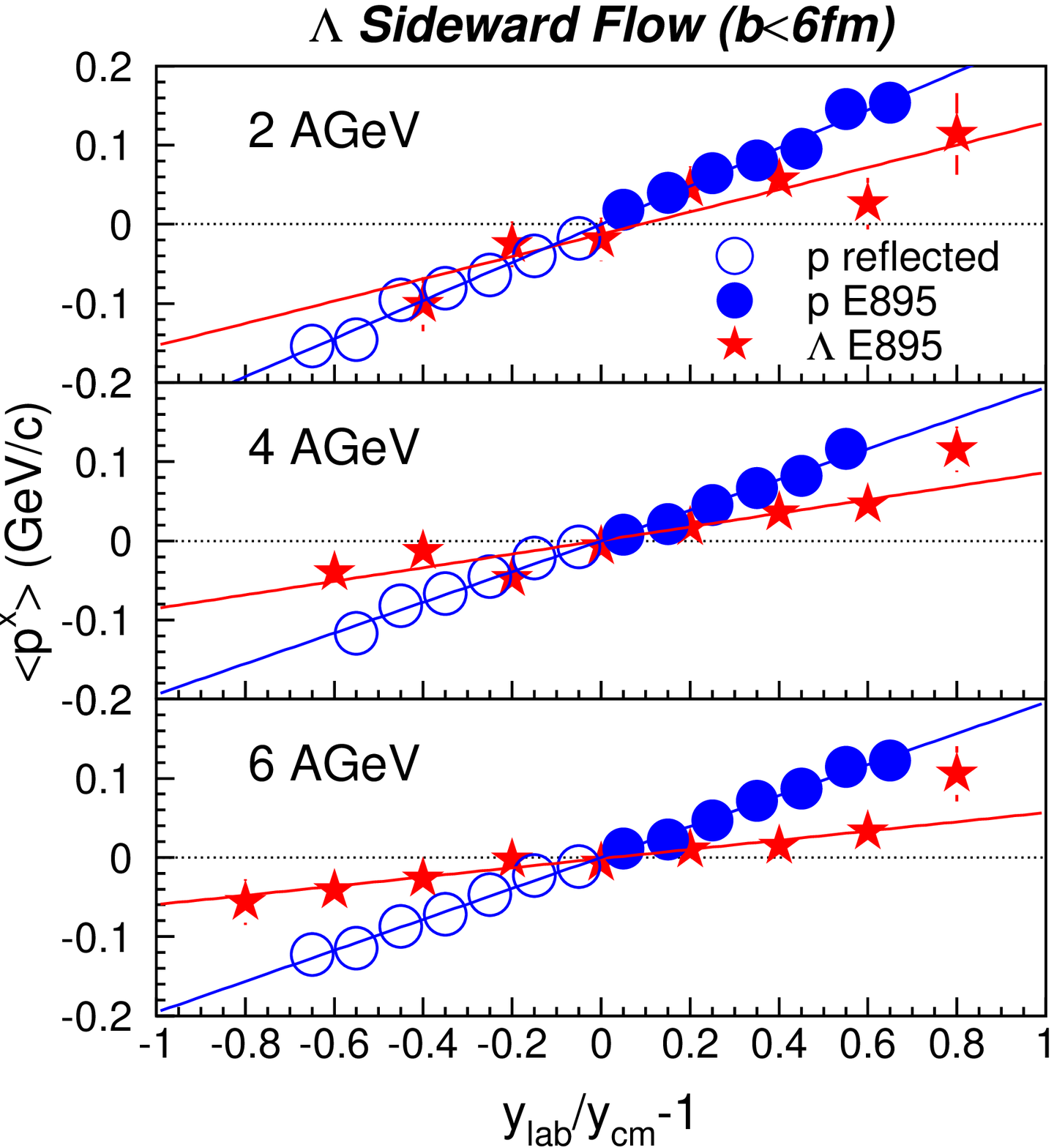,width=7.5cm}
}
\makebox[7.5cm][r]{
\epsfig{file=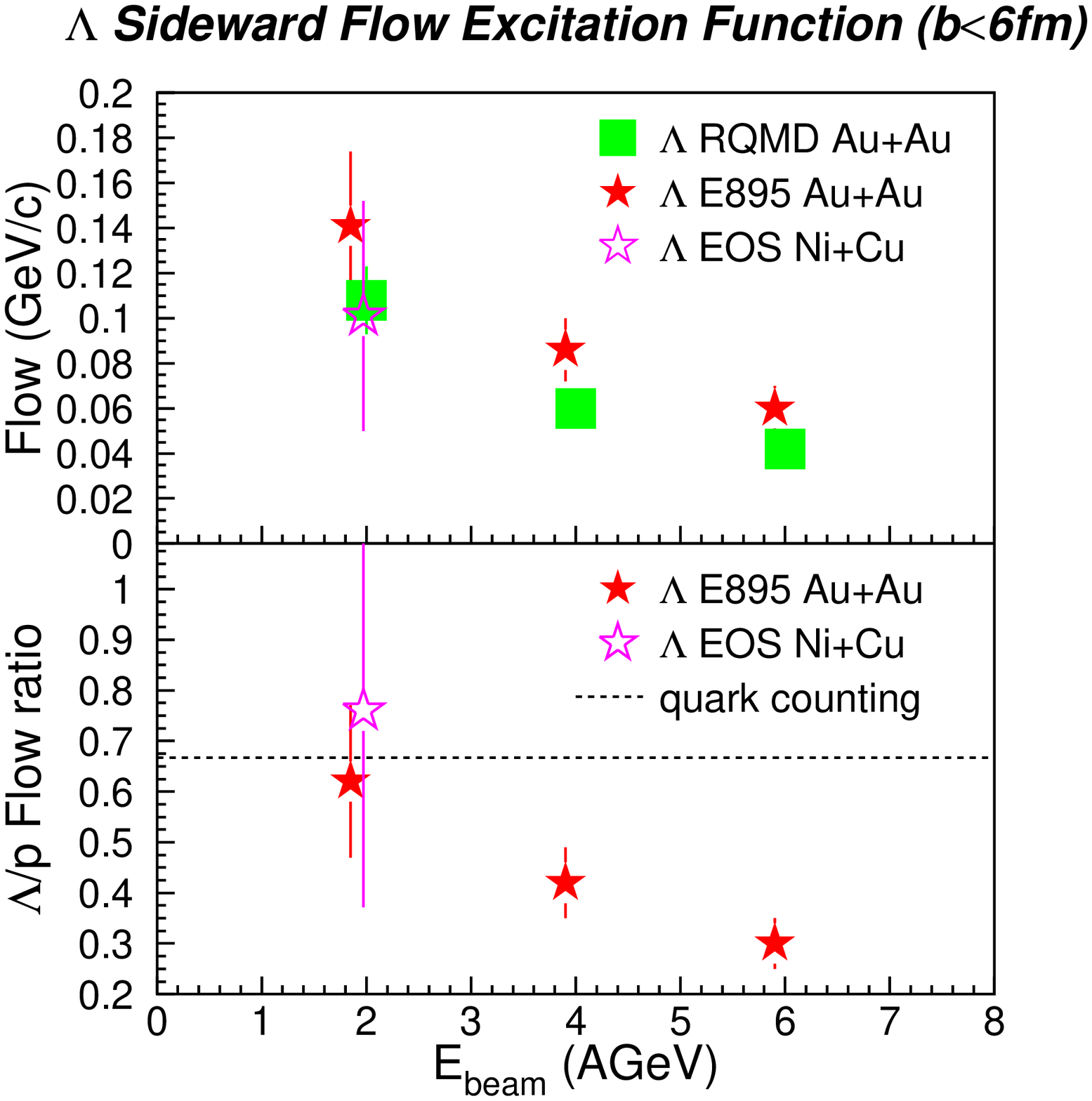,width=7.5cm}
}
}
\vspace{-10mm}
\caption[]{
\label{fig:lamflow}
\renewcommand{\baselinestretch}{1.} \em
$\Lambda$ Baryon sideward flow excitation function. The strength of the 
sideward flow is decreasing as a function of beam energy. With increasing
beam energy, a deviation from the simple quark counting rule is observed,
indicating the influence of the $\Lambda$ potential in dense matter.
}
\vspace{-5mm}
\end{figure}

The sideward flow analysis used the procedure outlined in \cite{Pinkenburg1999}
to reconstruct the reaction plane for each event with
the necessary corrections for the reaction plane dispersion.
Subsequently the projection of the momentum of each strange particle on its
associated reaction plane vector was obtained. The average momentum 
transfer ($\langle$p$^x$$\rangle$)
into the reaction plane for protons and $\Lambda$ Baryons is shown in 
Fig.\,\ref{fig:lamflow}. The $\Lambda$ Baryons exhibit a smaller sideward flow
than protons which where sampled from the same events in which a 
$\Lambda$ Baryon was reconstructed. This condition results in a bias of the 
proton sideward flow towards
more central events than 
reported in \cite{Heng1999}.
A comparison to a previous measurement
in the Ni+Cu System at 2.0 AGeV \cite{Justice1998} shows an increase of
the flow with system size similar to protons \cite{Chance1997}. 
If flow effects of $\Lambda$ Baryons are only
due to interactions of its non-strange quark constituents, one would expect
a 2/3 ratio of the strength of the $\Lambda$/proton 
flow \cite{Moszkowski1973} shown in
Fig.\,\ref{fig:lamflow}. This is indeed the case for the 2 AGeV data but
it deviates significantly at higher energies, indicating that there may be
additional contributions by the strange quark.

\begin{figure}[h]
\vspace{-5mm}
\centering\mbox{
\makebox[7.5cm][l]{
\epsfig{file=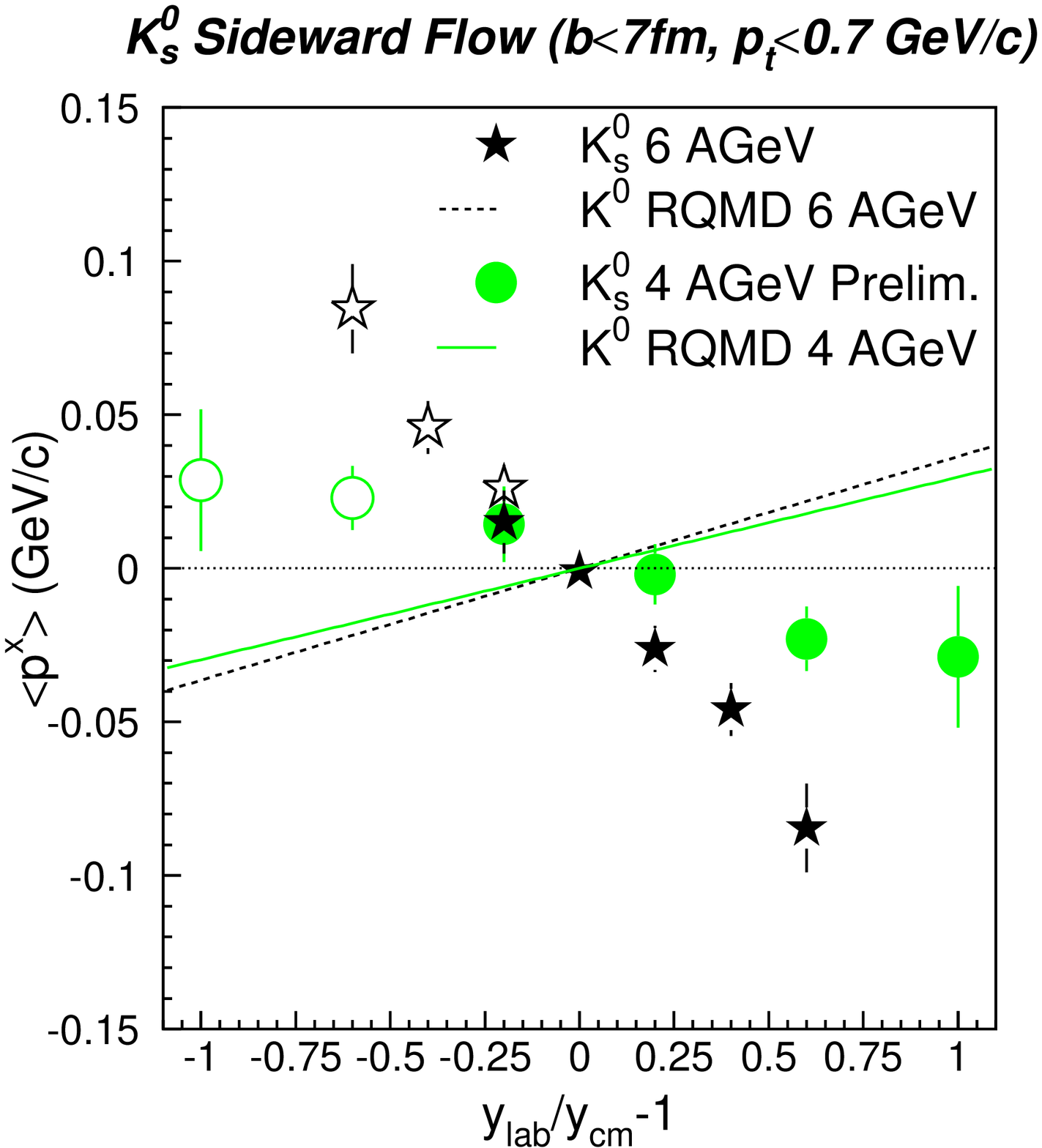,width=7.5cm}
}
\makebox[7.5cm][l]{
\epsfig{file=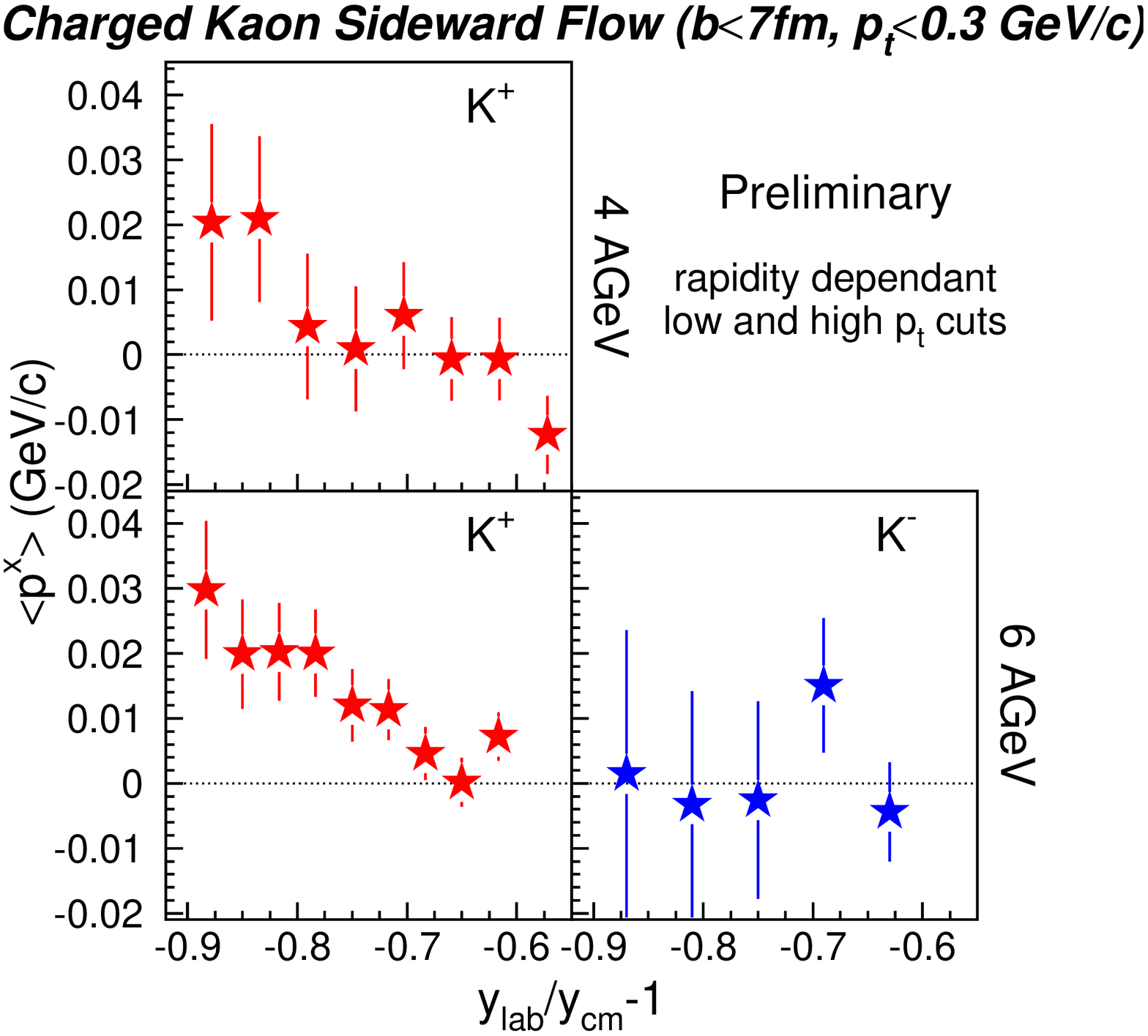,width=7.5cm}
}
}
\vspace{-1cm}
\caption[none]{
\label{fig:k0flow}
\renewcommand{\baselinestretch}{1.} \em
K$^0_s$ Mesons (open symbols represent reflected data) show a strong 
anti-flow signal which increases with beam energy.
The same behavior is qualitatively observed in preliminary results for positively charged Kaons while
the flow behavior of K$^-$ Mesons is compatible with no flow.
}
\vspace{-5mm}
\end{figure}

The K$^0_s$ Mesons depicted in Fig.\,\ref{fig:k0flow} exhibit a strong 
anti-flow
signal which increases strongly with beam energy. The RQMD model, which does 
not contain a kaon potential, can serve as a baseline for the effects of 
scattering. Its failure to reproduce even the sign of the flow is a strong hint
for the existance of a repulsive Kaon potential. 
Invoking a strong density
dependence of the Kaon potential, a calculation with ART is
able to reproduce the sign and the strength of the  K$^0_s$ Meson 
flow \cite{Pal2000}.
Since the K$^0_s$ Mesons at these energies originate mainly from 
K$^0$ Mesons --
according to RQMD the \={K}$^0$/K$^0$ ratio is smaller than 0.1 -- one 
would expect
a similar flow pattern for the K$^+$ Mesons. Indeed, even though the acceptance
for charged Kaons in E895 is very limited 
([y$_{Lab}$/y$_{c.m.}$-1]\,$<$-0.6) 
with a
rapidity dependent low and high p$_t$ cut, the K$^+$ Meson shows a clear 
anti-flow 
signal displayed in
Fig.\,\ref{fig:k0flow} which also increases with beam energy. 
The flow signal
of the K$^-$ Mesons is compatible with no flow.

The E895 collaboration has measured neutral strange particle production and 
flow at AGS energies. The results indicate that production yields are 
consistent with existing systematics and show no apparent dramatic increase
with beam energy. The transverse momentum analysis shows that $\Lambda$ Baryons
flow with protons but with smaller magnitude. K$^0$ and K$^+$ Mesons follow
an anti-flow pattern indicating a repulsive Kaon potential.

\end{document}